\documentclass[twocolumn,showpacs,preprintnumbers,prc]{revtex4-2}
\usepackage{graphicx,color}
\usepackage{float}
\usepackage{amsmath}
\usepackage{bm}
\usepackage[pdftex,colorlinks=true, linkcolor = blue, citecolor=blue,urlcolor=blue, bookmarksnumbered=true, bookmarksopen=true]{hyperref}
\begin{document}

\title{Variational approaches to constructing the many-body nuclear ground state for quantum computing}

\author{I. Stetcu}
\author{A. Baroni}
\author{J. Carlson}
\affiliation{
  Los Alamos National Laboratory, Theoretical Division, Los Alamos, New Mexico 87545, USA}
  
\date{\today}
\preprint{LA-UR-21-29364}

\begin{abstract}

   We explore the preparation of specific nuclear states on gate-based quantum hardware using variational algorithms.  
   Large scale classical diagonalization of the nuclear shell model have reached sizes of $10^9 - 10^{10}$ basis states, but are still severely limited by computational resources. Quantum computing can, in principle, solve such systems exactly with exponentially fewer resources than classical computing. Exact solutions for large systems require many qubits and large gate depth, but variational approaches can effectively limit the required gate depth.
   
We use the unitary coupled cluster approach to construct approximations of the ground-state vectors, later to be used in dynamics calculations. The testing ground is the phenomenological shell model space, which allows us to mimic the complexity of the inter-nucleon interactions. We find that often one needs to minimize over a large number of parameters, using a large number of entanglements that makes challenging the application on existing hardware. Prospects for rapid improvements with more capable hardware are, however, very encouraging.
\end{abstract}
\maketitle

\section{Introduction}

Quantum computing holds the promise of exact solutions for specific quantum states and dynamics~\cite{Feynman1982,Lloyd1996,Farhi2000}, which in nuclear systems~\cite{Dumitrescu2018} would revolutionize our ability to understand and nuclei as probes of fundamental physics, create medical isotopes, and diagnose complex astrophysical and terrestrial environments.
   While this ultimate goal awaits the development of quantum hardware with more qubits and smaller error rates, we can begin to examine potential near-term algorithms on quantum hardware which help us move toward these ultimate goals.
   One clear important area of investigation is variational approaches for preparing specific quantum states.
   
   In this paper we first examine required quantum resources in number of qubits and circuit depths for quantum problems. We then comment on various approaches to solving the nuclear many-body problem in the shell model, traditionally formulated in terms of harmonic oscillator single particle states, and lattice approaches. In the more standard gate-based quantum hardware, the number of qubits required is largely determined by the single-particle space considered. The circuit depth for exact solutions is largely determined by spectral properties of the nuclear Hamiltonian.  It is here that variational approaches are helpful in that they can produce an accurate, though approximate, initial state that be made exact with
   dynamical quantum algorithms and/or used in calculations of nuclear transition rates, response and more general scattering processes.
   
The bulk of the paper is devoted to the use of the unitary coupled cluster approach to construct approximations of the ground-state vectors. The testing ground is the phenomenological shell model space, which allows us to mimic the complexity of the inter-nucleon interactions. 
We find that it is important to optimize a large number of variational parameters in this approach, which is challenging on current hardware.  Improved quantum hardware will lead to rapid advances in the size of problems that can be handled, however. In addition, recent algorithmic improvements in VQE algorithms may further reduce the required gate depth~\cite{Tang2021}.

\section{Theoretical framework}

The nuclear many-body problem is very similar to other strongly-correlated quantum many-body problems including 
many-electron systems in atomic and molecular physics, properties of bulk hydrogen and helium in the cores of gaseous planets, and cold atom and molecular systems including unitary fermions.   Though nuclei are strongly-correlated, they do have a substantial mean field which governs many low-energy properties including binding energies, radii, and electromagnetic transition rates. These properties have 
traditionally been explored in the nuclear shell model
through large-basis diagonalizations with phenomenological two- and sometimes three-nucleon interactions.
In such calculations the single-particle states are typically treated as solutions of a harmonic oscillator to enable a clean separation between relative and center-of-mass coordinates.
While the single-particle model spaces are typically rather constrained, of order 10-100 states, the total dimensionality of the problem grows extremely rapidly with particle number.
Even with restrictions upon total oscillator energy in the many-body states, the full dimensionality of the problem can easily exceed the capacity of the largest classical computers.

For quantum many-body problems, it is natural to consider the standard Jordan-Wigner (JW) \cite{JW} or Bravi-Kitayev (BK) \cite{BRAVYI2002210} encodings of the quantum many-body problem.  Though the interaction is more complicated than, for example, Coulomb interactions in atomic and molecular systems, it is 2-local (or 3-local with three-body interactions) making the encoding relatively efficient.  A fault-tolerant quantum computer with 50-100
qubits could easily exceed the capabilities of present-day classical computers. In addition to ground-state properties quantum computers can be used to calculate dynamical properties including response functions and hadronic scattering.

Treating 'realistic' nucleon-nucleon interactions fit to nucleon-nucleon scattering data is more difficult. In this case the single-particle basis has to be much larger.  For example a cubic lattice with a total length L in each dimension of 10 fm and a 1 fm lattice spacing would require 1000 qbits with BK encodings.
For small particle numbers first-quantized approaches to the encoding, where the number of qubits scale logarithmically with the number of basis states would be more efficient.

Beyond just the number of qbits, the circuit depth is very important for near-term applications.
Here variational approaches can be important in reducing the circuit depth, producing accurate approximate solutions in relatively few gates appears feasible in typical shell-model applications, as we discuss below.

\subsection{Model space and interactions}

In general, nucleons interact via two- and higher body interactions, with the dominant contribution coming from two-nucleon interactions. Hence, in this paper, we limit ourselves to two-body interactions. In addition, in order to be able to run some of our simulations on quantum hardware, we have decided to restrict ourselves to rather small model spaces, where we can still use some ``realistic" two-body interactions. For this reason, we have adopted the phenomenological shell model interactions derived at a time when the computational power was severely limited. In this model, one assumes that only a few (valence) nucleons interact, while the interaction with the inert (core) nucleons is approximated via a diagonal one-body term (single-particle energies). Thus, the Hamiltonian for such a model writes

\begin{equation}
    H=\sum_{i=1,N_s} \varepsilon_{i} a^\dagger_i a_i +\frac{1}{2} \sum_{ij,kl} V_{ij,kl} a_i^\dagger a_j^\dagger a_l a_k,
    \label{eq:hamSM}
\end{equation}
where $N_s$ is the number of states in the model, $a_i^\dagger$, $a_i$ are the creation and annihilation operators for state $i$, and $\varepsilon_i$ and $V_{ijkl}$ are the single particles and two-body matrix elements, that are fitted to reproduce energy spectra for a limited number of nuclei. However, the method presented here should be general enough to be applied in other approaches, for example the no-core shell model. Thus, this non-trivial problem has the advantage that while it is defined in a small model space (\textit{i.e.,} requires a relatively small number of qubits), it has a similar complexity to the more realistic inter-nucleon interactions. In the present investigation, we use the Cohen-Kurath interaction \cite{COHEN19651} in the $p$ shell model space, which includes six states for protons and six for neutrons and assumes a inert $^4$He core, and the ``universal sd"  Wildental interaction \cite{WILDENTHAL19845,PhysRevC.74.034315} in the sd model space, with 12 states available for protons and 12 for neutrons and a $^{16}$O inert core. In most of the calculations, we only include neutrons in order to keep the required number of qubits small enough for simulations. We consider only one case in which we treat two protons and two neutrons in the 0p shell.

In this paper we use unitary coupled cluster ansatz to construct a correlated state. This requires calculate the Hartree-Fock (HF) solution, which determines the occupied and unoccupied states. We use the code \texttt{SHERPA} to compute the mean-field solution in this basis \cite{PhysRevC.66.034301}. Because the rotational invariance is in general broken in such an approach, usual constrains like fixing the total angular momentum projection when constructing basis states can no longer be imposed, which in turn produces an increased number of configurations that are required to obtain a correlated state. While this is not ideal, by definition a quantum computer is able to handle such an increased workload. In the case where the HF solution is spherical, we can significantly limit the number of configuration that we include in the simulations, without affecting the quality of the solution, by allowing only two-particle, two-hole configurations that have zero total projection of the angular momentum on the $z$ axis. Note that while \texttt{SHERPA} can handle an odd number of  protons and neutrons, we restrict our investigations to even-even systems.

A consequence of using the HF solution is that the interaction in Eq. (\ref{eq:hamSM}) will have to be transformed to the HF basis, where the creation and annihilation operators describe deformed single particle states. In general, this increases the number of terms in the interaction, and is particle-number dependent, just like the mean field.

The shell model Hamiltonian is naturally given in second quantization, hence JW \cite{JW} or BK \cite{BRAVYI2002210} mappings of the states into qubits are an excellent match for a relatively straight-forward implementation. We have implemented both mappings and also considered the first quantization encoding, but we show in the current paper exclusively results with the Jordan-Wigner mapping. In particular, in our implementation of the first quantization approach, the Hamiltonian was mapped into considerably more Pauli terms, in addition to the challenges posed by the antisymmetrization of a large number of particles. 
The challenge posed by antisymmetrization can be circumvent using a recently developed quantum algorithm ~\cite{Berry2018a} that allows to antisymmetrize $\eta$ particles over  $N$ single-particle basis function with a gate complexity of $O(\log^c\eta\log\log N)$ and a circuit size of $O(\eta\log^c\eta\log N)$ and the value of $c$ depends on the choice of the specific algorithm. 
We remind the reader here that the problem of estimating the ground-state energy of a quantum many-body system is of great importance for quantum simulations. A series of non-variational approaches has been proposed that make use of Quantum Phase Estimation Ref.~\cite{Kitaev1995} and its improvements~\cite{Poulin2009,Ge2018,Lin2021a,PhysRevLett.127.040505}.  The best known algorithm has a scaling depending on the spectral gap $\Delta$, the overlap of the trial state with the ground state, and in general the probability of preparing a state with precision $\epsilon$ and probability $(1-P)$ leads to a gate depth that scales as (corollary 9 of Ref.~\cite{Lin2021a})
\begin{eqnarray}
{\cal O}\left( \frac{1}{\gamma\Delta}\left[ \log\frac{\alpha}{\Delta}\log\frac{1}{\gamma}\log\frac{\log\alpha/\Delta}{P}+\log\frac{1}{\epsilon}\right]\right)
\end{eqnarray}
and of course for cases in which $\Delta$ and (or) $\gamma$ are small becomes unfeasible. 
Additionally, unitary evolution on quantum computers producing the equivalent of Lanczos, or imaginary-time algotithms similar to quantum Monte Carlo have been introduced~\cite{Motta:2020,Baker:2021}. Similar to Quantum Phase Estimation, these algorithms also produce exact answers in principle, but require a larger number of gates and higher fidelity.
For current generation computers the use of variational algorithms is preferred~\cite{Cerezo2021b}, and may be very valuable even in the future to 
prepare accurate starting points for more exact methods.

\subsection{The unitary coupled cluster ansatz}
Probably the most widely used approach to generating a correlated ground-state solution from the HF state, also suitable for implementation on quantum hardware, is the unitary couple cluster (UCC) method. Formally, this reduces using an anti-Hermitian unitary transformation $U(\vec \theta)$  so that the trial state
\begin{equation}
    |\Psi(\vec \theta)\rangle=\exp\left(U(\vec \theta)\right)|\Psi_0\rangle
    \label{eq:anzatz}
\end{equation}
minimizes the energy by adjusting the parameters $\vec \theta$. In Eq. (\ref{eq:anzatz}), $|\Psi_0\rangle$ stands for the HF solution. In UCC, the anti-Hermitian operator $U(\vec \theta)$ is assumed to have a simple one- or more particle-hole configuration form
\begin{eqnarray}
\lefteqn{
    U(\vec \theta)=\sum_{i,m} \theta_{im} \left( a^\dagger_ia_m- a^\dagger_m a_i \right)} \nonumber \\
    && +\sum_{i<j;m<n}\theta_{ij;mn}\left( a^\dagger_ia^\dagger_ja_n a_m- a^\dagger_m a^\dagger_n a_j a_i \right)+\cdots,
    \label{eq:U}
\end{eqnarray}
\iffalse
{\color{blue}What are the symmetries of the coefficients $\theta_{ijmn}$? In chemistry there are symmetries see  for  example Ref.~\cite{Motta2021a} after Eq. 7, using such symmetries approximate implementations of the UCC ansatz have been proposed \cite{Motta2021a}}  
\fi
where the sums run over occupied ($i,j$) and unoccupied ($m,n$) states. In many applications, restriction to two-particle, two-hole configurations gives very good approximations to the exact solution, but there are cases where higher-order terms have to be considered to improve the quality of the solution.

\begin{figure}
    \centering
    \includegraphics[scale=0.35]{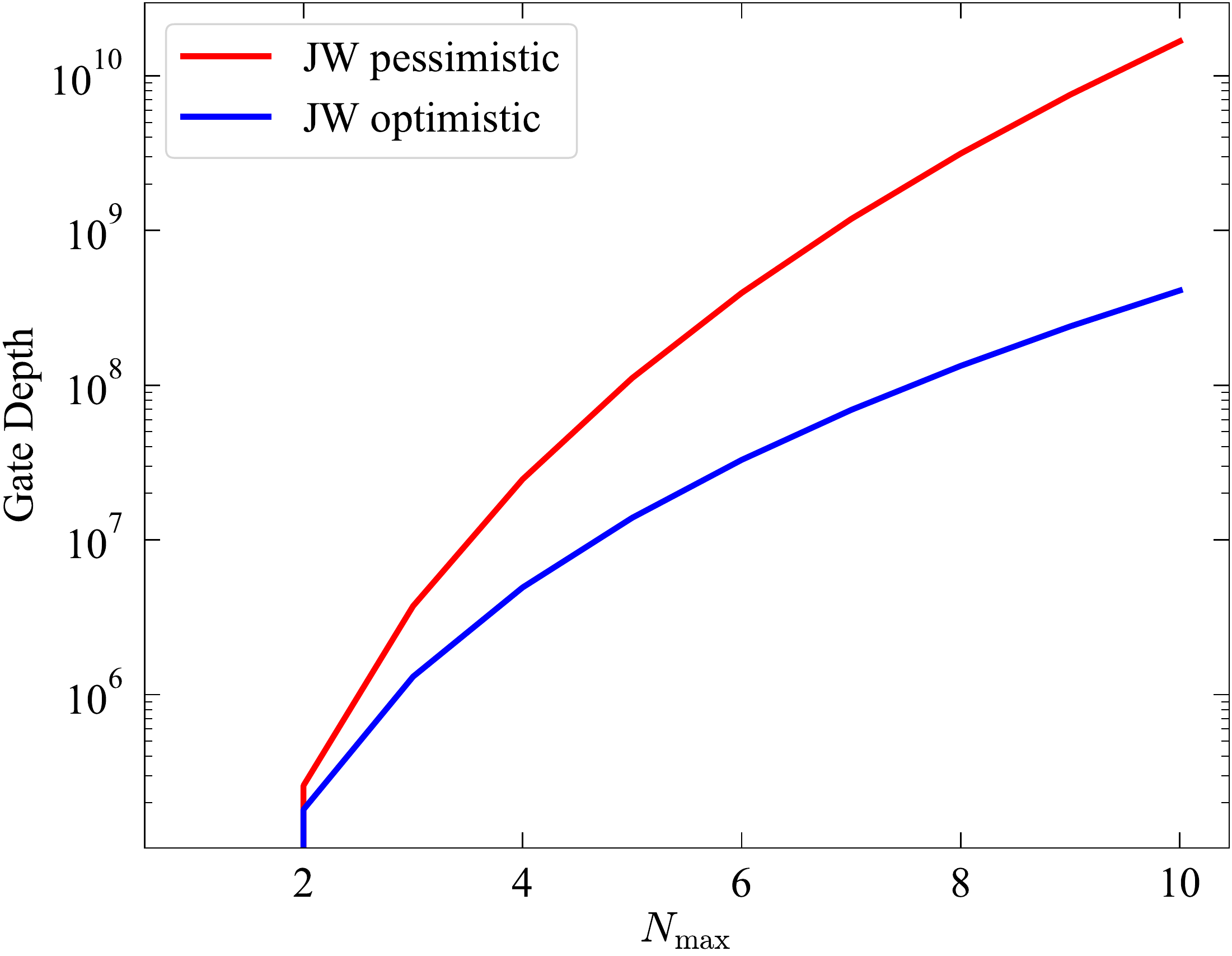}
    \caption{Gate depth as a function of the shell model parameter $N_{\rm max} $ for $^{16}$O for three different UCC methods. The current biggest classical calculation uses $N_{\rm max}=8$ in the single-particle basis, but also impose a cutoff on the total many-body excitations in the harmonic oscillator basis to reduce the basis size. }
    \label{fig:gate_depth}
\end{figure}

In addition to the truncation to a tractable number of configurations, the Trotter approximation is often invoked as a source of errors because the implementation of the exponential of the sum of operators is non trivial. While in some cases improving the Trotter approximation helped \cite{TrotterDef}, for the many-body systems considered in this paper, the impact of such an improvement was negligible. Note that because the terms corresponding to each parameter $\theta$ in Eq. (\ref{eq:U}) conserves the number of particles, the Trotter approximation does not introduce errors that would break the particle number. We report in Fig.~\ref{fig:gate_depth} the circuit depth as a function of the $N_{\rm max}$ in the shell model. 
As we can see the parallelization schemes such as fermionic swap can reduced the number of layers needed by some order of magnitudes. The current gate depth remain, however, unfeasible for the expected first generation fault tolerant quantum computers.
We analyze now the scaling of the UCCSD ansatz.
Our discussion follows the one reported in Ref.~\cite{Cao2018}, adjusted for the problem at hand. For the case of a first order Trotterized UCCD ansatz the total number of parameters (equivalent to the number of doubles) is
\begin{eqnarray}
\lefteqn{N_{\rm param.}=N_pN_n(N-N_p)(N-N_n)}
\nonumber \\
  && +\binom{N-N_p}{2}\times \binom{N_p}{2}+\binom{N-N_n}{2}\times \binom{N_n}{2},
\end{eqnarray}
with $N_p$ and $N_n$ the number of protons and neutrons, respectively, and $N$ the number of single particle states included in the calculation (taken for simplicity equal for both protons and neutrons). 
and we consider  the case  for which $N\gg N_{n,p}$ that leads to a number of parameters that scales as  ${\cal O}(N^2(N_p+N_n)^2)$. It is convenient to denote with  $f$ the gate depth of a single  term associated of the  UCCD ansatz, therefore the scaling of the gate depth is ${\cal O}(N^2(N_p+N_n)^2\,\,f)$. In the worst case scenario, for the case of Jordan-Wigner and Bravyi-Kitaev transformations we have, respectively, $f\in{\cal O}(N)$  and $f\in{\cal O}(\log(N))$. These depths are for serial executions  and recent work in  quantum chemistry has made progress using parallelization techniques \cite{Kivlichan17}.  In particular a significant reduction in the gate depth can be obtained using a fermion swap network, as discussed in Ref.~\cite{Gorman2019}, for the nuclear problem at hand this means to have $O(N^2(N_p+N_n)^2)$ layers. 
\iffalse
We now give a brief overview of the fermion swap network for the case of the UCCD ansatz. 
Things can be further improved using the low rank approximation reported in Ref.~\cite{Motta2021a} that leads to an overall scaling of $O(N^4)$.

{\color{red}Before moving to the actual implementation it would be important discuss how the UCC does for systems that can be run on an actual laptop, see for example 
\url{https://arxiv.org/pdf/2005.08451.pdf} figure 3. I would also be interesting to compare the Trotter UCCSD ansatz along with the QCCSD and other variations.
{\color{red} Expand and discuss briefly what is a fermion swap network, cite the neutrino paper as an early application.} 
}
\fi

\subsection{Elementary operator mappings}

There are two types of mappings that allow simulations of physical systems on quantum hardware. The most straight forward approach is the second quantization mapping, in which elementary every fermionic state associated with a distinguishable qubit. In such approaches the creation and annihilation operators for fermionic states are written in terms of Pauli operators acting on different qubits. For example, in the Jordan-Wigner (JW) approach, the creation and annihilation operators associated with state $i$ is mathematically formulated as \cite{JW}
\begin{eqnarray}
a^\dagger_i = && \frac{1}{2} \left(\prod_{j=0}^{i-1}-Z_j\right) (X_i-iY_i),\\
a_i = && \frac{1}{2}\left(\prod_{j=0}^{i-1}-Z_j \right)(X_i+iY_i). 
\end{eqnarray}
Here the $X_q$, $Y_q$ and $Z_q$ are the Pauli matrices acting on qubit $q$. Hence, all many-body operators, including the Hamiltonian, can be mapped into sets of Pauli operators acting on different qubits. The occupation number in the JW mapping is stored in the $|0\rangle$ and $|1\rangle$ state of the qubit, corresponding to unnoccupied and occupied state respectively. While simple, this method has the disadvantage that in order to represent state $i$ one must include $i-1$ operators acting on the previous qubits. For an early application of this method in nuclear physics, see Ref. \cite{Dumitrescu2018}.

An alternative to the JW mapping is the parity representation, in which the $q^\mathrm{th}$ qubit stores the sum of the parity of the the first $q$ mode \cite{Seeley:2012,RevModPhys.92.015003}. This mapping has not yet been proven very useful in applications to quantum chemistry \cite{RevModPhys.92.015003}.

The Bravyi-Kitaev (BK) mapping \cite{BRAVYI2002210} has been proved to be as efficient, and in many cases considerably more efficient in quantum chemistry calculations of ground states of molecular systems \cite{Tranter:2015,Tranter:2018}. In this approach, the qubits store partial sums of occupation numbers, but require the number of states to be powers of 2. However, an operator that has a weight $\mathcal{O}(M)$ in the Jordan-Wigner encoding would only have  $\mathcal{O}(\log_2(M))$ in the BK mapping.

The second quantization mappings, like those briefly described above, have the disadvantage that the number of qubits scales with the number of single particle states, which grows very fast in realistic nuclear physics problems. In a first quantization approach, a better scaling can be achieved, at a cost: the anti-symmetrization, which naturally enters in the JW mapping, must be implemented explicitly for a relatively large number of particles \cite{AbramsLloyd1997}.
However, in the long run, with efficient anti-symmetrization methods, this mapping would be better suited for future applications as it has better scaling than JW or BK mappings \cite{RevModPhys.92.015003}, especially in applications where the problem is discretized on large lattices.

\begin{figure*}[t]
 %   \centering
    \includegraphics[scale=1]{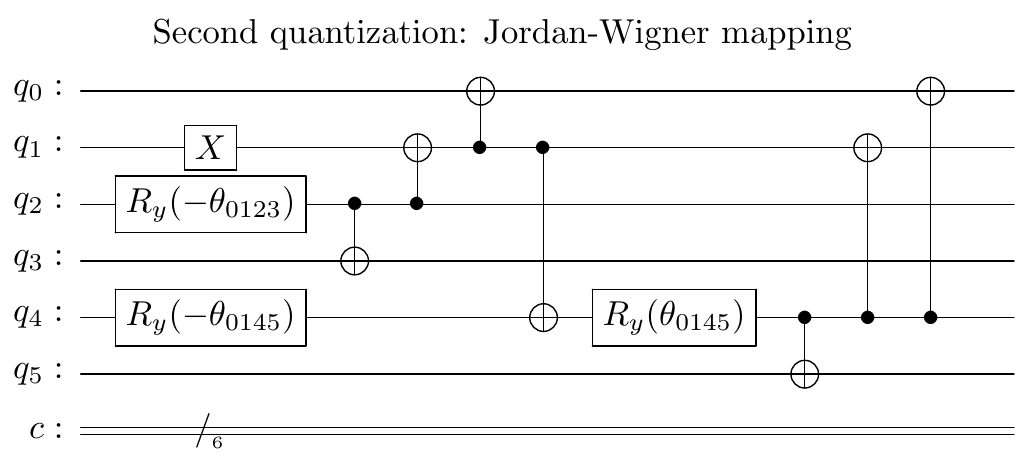}
    \includegraphics[scale=1]{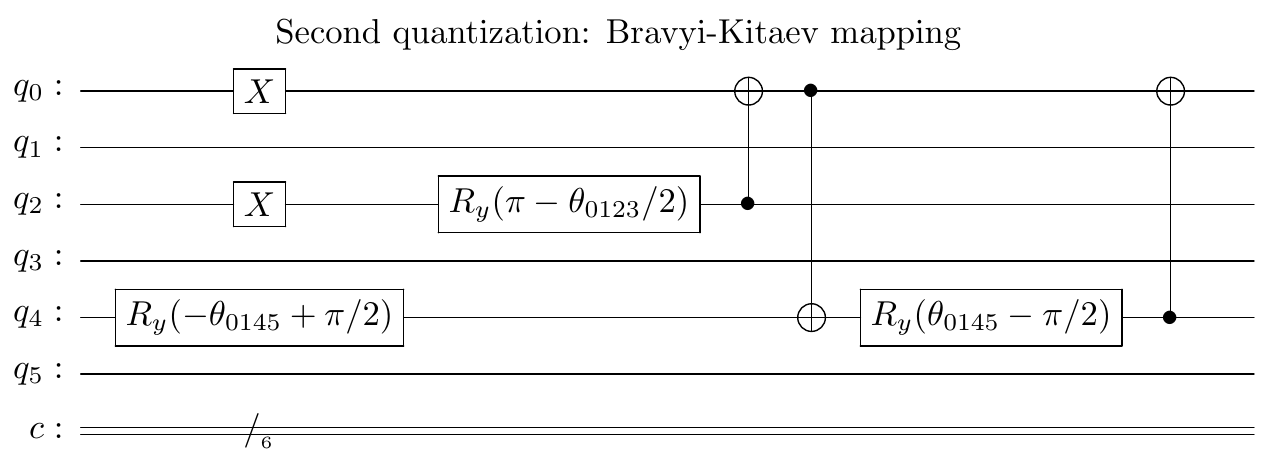}
    \includegraphics[scale=1]{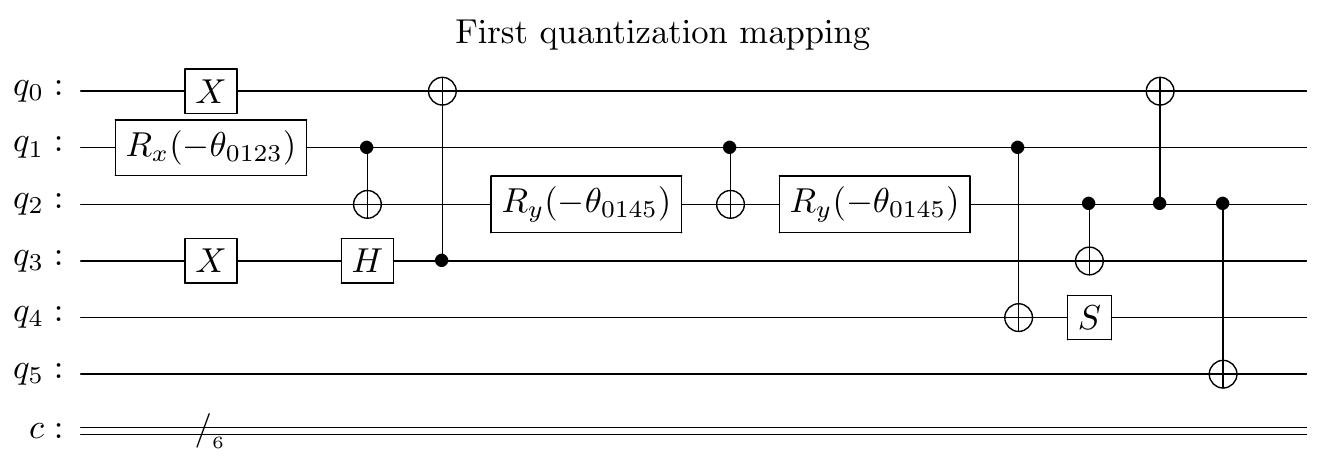}
    \caption{Circuits for including a limited number of two-particle, two-hole configurations on top of the HF solution for two particles in 6 states, with $R_y(\theta)=\exp(i\theta Y/2)$. All three mappings are shown in this figure: Jordan-Wigner (upper panel), Bravyi-Kitaev (middle panel) and first quantization (lower panel). These circuits can be extended to more single particle states and select extra two-particle, two-hole configurations. When $\theta_{0 123}=\theta_{0 145}=0$, the Hartree-Fock state is recovered.}
    \label{fig:circ-2p2h-JW}
\end{figure*}

A relatively simple circuit can be constructed for including two-particle two-hole configurations on top of the HF solution for two particles (\textit{e.g.,} neutrons) in the $p$ shell, as shown in Fig.~\ref{fig:circ-2p2h-JW} for all three mappings discussed above. In this case only the two-particle, two-hole configurations promoting particles from states $0,1$ to $2,3$ and $4,5$ respectively have significant contributions. The circuit in the BK mapping in particular can be further reduced by removing $q_1$, $q_3$ and $q_5$, and analytically computing all expectation values on these qubits. The number of terms in the Hamiltonian mapping is reduced from more than 170 to 13. In Sec. \ref{sec:results}, we use this reduced circuit to minimize the energy. It is not likely that such a reduction will be possible for more complicated configurations, but in some particular cases one could possibly make some calculations more efficient by performing similar removal of select qubits. 

In a more general case, for more particles or where the number of two-particle, two-hole configurations that are important is large, the complexity of the circuit increases rapidly. The circuits represented in Fig. \ref{fig:circ-2p2h-JW} are not universal, and it can only be applied to two particles and only because we did not include additional excitations (\textit{e.g.,} promoting particles from states from $0,1$ to $2,4$ on top of the correlations included already in the circuits).
%{\color{red} We should probably report  the full expression, at least for one case of the UCCSD ansatz for the analogue circuits of the one used in fig.2 and then comment on the excessive gate depth, even using identities, and say that the circuit in Fig.2 are a way to  get around the problem for runs on current hardware, although they are not easily generalizable.}

For more general circuits that can handle a larger number of particles and excitations, we have used the work in Refs. \cite{SD_excitations0,SD_excitations}  to implement particle-conserving one- and two-body correlations. Because we have included four-particles, four-holes in a very limited number of tests, we have used a less efficient circuit for testing purposes only. It consists of a ``brute-force" approach, in which we expand the set of four creation and four annihilation operators in Pauli strings, and take the exponential of each Pauli product, as they commute, similar to Eqs. (A6)-(A7) in Ref. \cite{PhysRevA.98.022322} extended to higher particle-hole correlations. This results in a large number of CNOT entanglements that are out of reach for the hardware available today.

\subsection{Entanglement}

One of the main questions is how entangled are the states that need to be prepared, as the complexity of the circuits required to produce such states depends on this entanglement. It is difficult \textit{a priori} to estimate the degree of entanglement. However, because the systems that we consider in this paper have a numerical solution, we can investigate this question in the context of entanglement entropy and mutual information. 

The density matrix for a prepared state $|\Psi\rangle$ (in this case ground state) is written as
\begin{equation}
    \rho=|\Psi\rangle\langle \Psi |.
    \label{eq:rhofull}
\end{equation}
As discussed in many references before us, one can always consider a partitioning of the basis states in two subsystems and trace the density matrix in Eq. (\ref{eq:rhofull}) over one of the subsystems, obtaining a reduced density matrix. The reduced density matrix can be diagonalized, with eigenvalues $\rho_i$, where $i$ runs from 1 to $2^{n_s}$, with $n_s$ the number of single particle states not included in the trace. The von Neuman entropy is calculated as 
\begin{equation}
    S=-\sum_i \rho_i \ln\rho_i.
\end{equation}
This entanglement entropy has two extremes: 0, when the subsystems are decoupled, and $n_s\ln(2)$, when the subsystems are maximally entangled.

\begin{figure*}[t]
    \centering
    \includegraphics[width=\columnwidth]{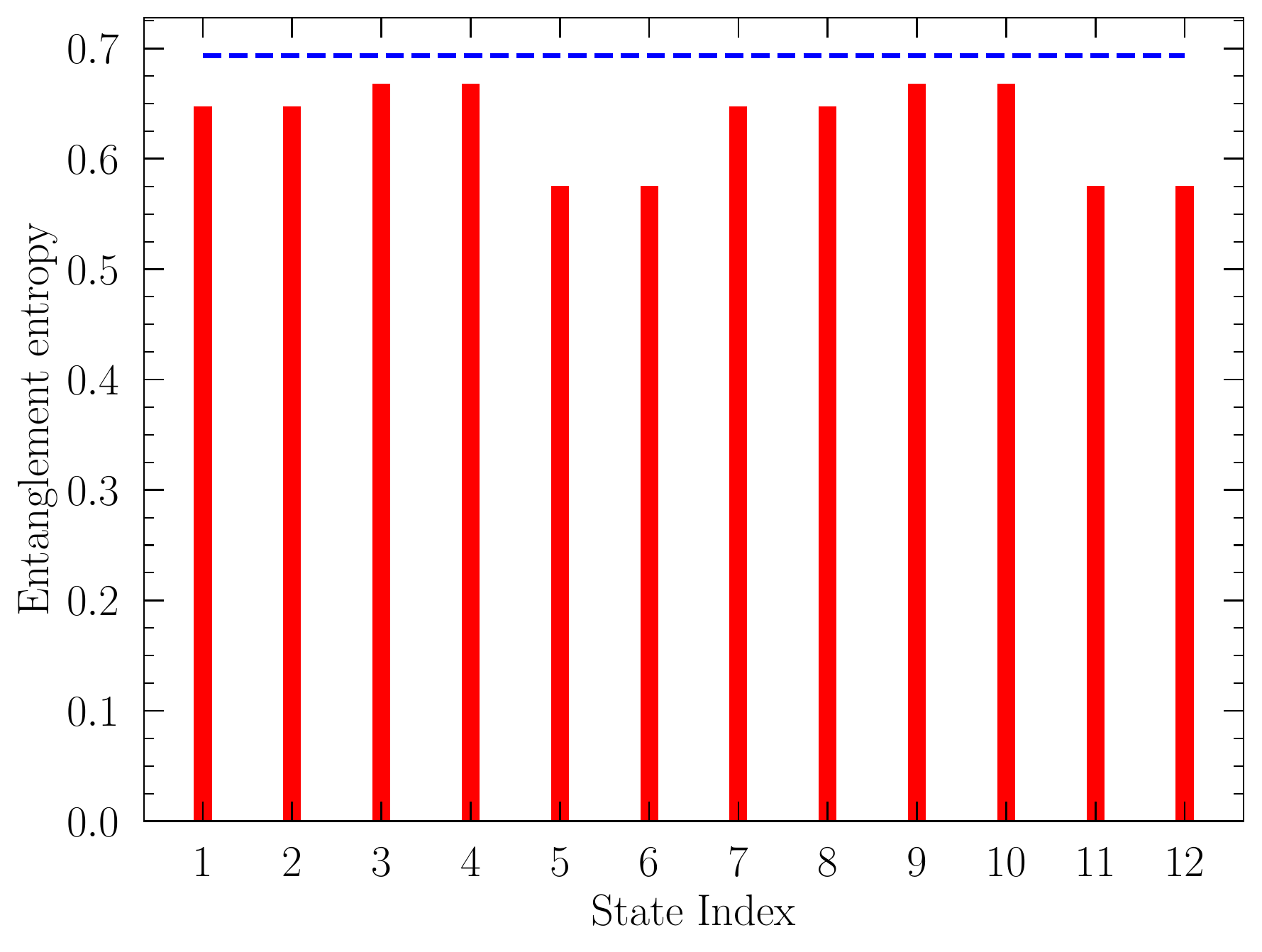}
    \includegraphics[width=\columnwidth]{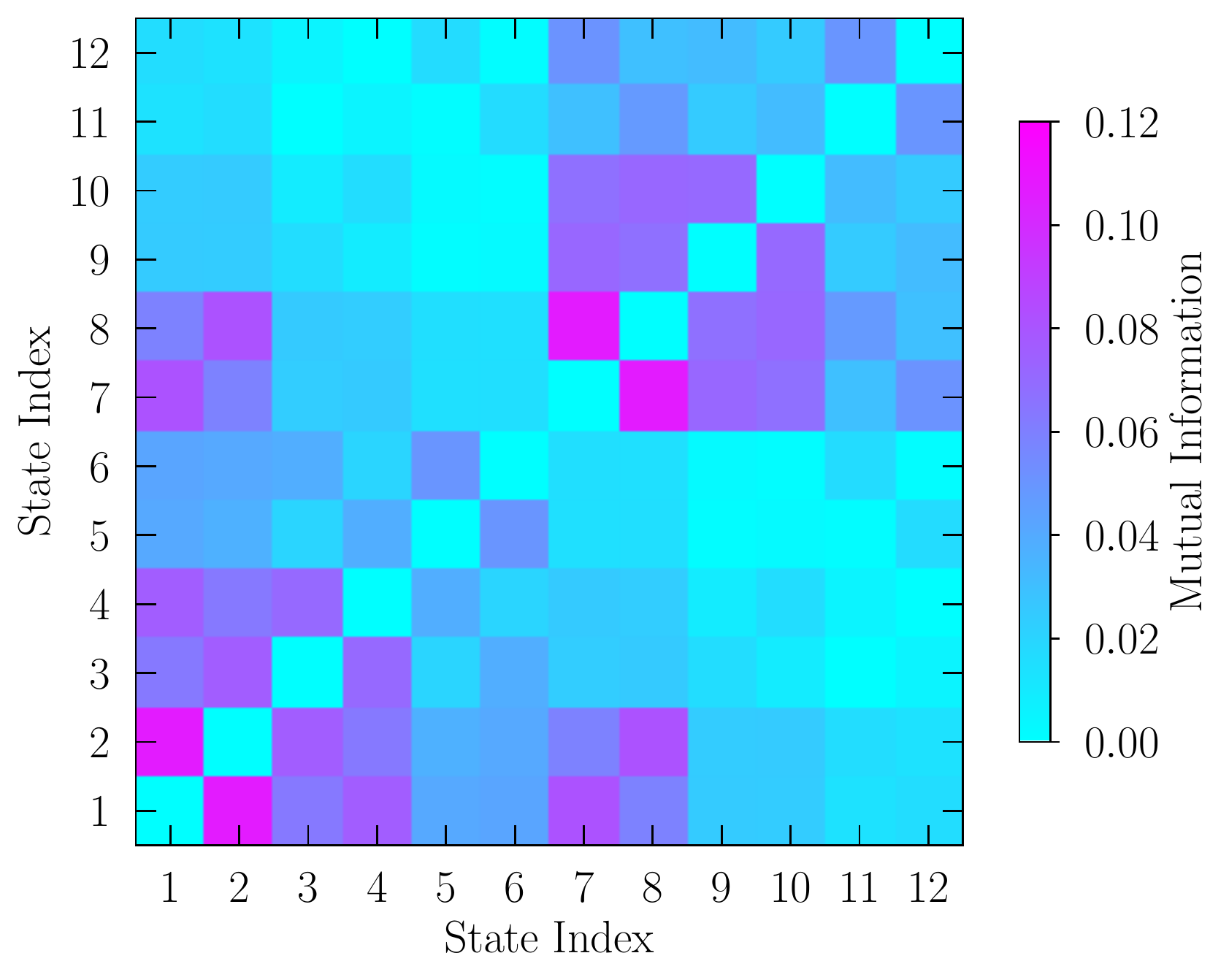}
    \caption{Left panel: the s.p. entanglement entropy (red bars) compared against the maximal entanglement value (dashed blue line). Right panel: the mutual information for the HF s.p. states. In both cases, we have used the s.p. HF states of $^8$Be in the p shell. States 1 through 6 are for neutrons and 7 through 12 are for protons.}
    \label{fig:MutualInformation}
\end{figure*}

In the following, we consider the entanglement of each HF state, so that $n_s=1$, and in order to construct the reduced density matrix, we will trace over $N-1$ states. Furthermore, we define like in quantum chemistry and nuclear physics studies \cite{RISSLER2006519,Robin:2021} the mutual information between two states $\alpha$, $\beta$ as

\begin{equation}
    I_{\alpha\beta}=\frac{1}{2}\left( S_\alpha+S_\beta-S_{\alpha\beta}\right) (1-\delta_{\alpha\beta}),
\end{equation}
which is a quantitative measurement of the correlations between the two orbitals. This quantity is zero if there two orbitals are not entangled. In Fig. \ref{fig:MutualInformation} we plot the entanglement entropy of the HF s.p. states and the mutual interaction among all (proton and neutron) HF s.p. states, for a calculation of $^{8}$B in the p shell. The s.p. entanglement entropy is almost maximal, consequence of the high-correlations in the small model space considered here. All quantities were extracted from the exact calculations, using the exact solution obtained by diagonalizing a matrix with dimension 225. It can be seen immediately that the entanglement between different protons and neutrons orbitals is quite strong, particularly between the lowest proton and neutron states, presumably because the model space is so restricted. In phenomenological approaches like the shell model, one treats mostly states around the Fermi surface, which are expected to be highly correlated.

\section{Results}
\label{sec:results}

We have implemented the parametrized quantum circuits reported in Fig.~\ref{fig:circ-2p2h-JW} on the available simulators and, in some cases, on current quantum hardware. The classical minimization procedure that enters in VQE has been carried out  using different versions of gradient descent, currently implemented in the method  \texttt{minimize()} from the \texttt{scipy} module.
\iffalse
We have used different approaches to minimization, depending on whether we or not we were able to run on quantum hardware or simulators on cloud or  local computers. For local simulators, we use IBM's \texttt{QISKIT} combined with the minimization routine \texttt{minimize()} from the \texttt{scipy} package.
\fi

\subsection{Simulator results}
The simplest problem we can run on 6 qubits is two neutrons in the $p$ shell, which corresponds to $^6$He. This is not to say that our solution properly describes the wave function for this weakly-bound nucleus in such a small space. As noted in the previous section, we obtain a very good approximation of the ground-state energy if we consider only two parameters, as shown in Table \ref{table:results}. This is not necessarily surprising, since in a shell model implementation there are only 3 states with the total projection of $J_z$ summing up to zero. 

We have extended the calculations to other nuclei as well, $^8$Be (two protons and two neutrons in the $p$ shell) and $^{20,22}$O (four and six neutrons respectively in the $sd$ shell), with all the results summarized in Table \ref{table:results}. For $^8$Be system, we show the results when we include only two-particle, two-hole configurations for the same type of particles (protons or neutrons), adding two-particles, two holes configurations that include proton and neutron excitations at the same time (marked by an asterisk), and two protons and two neutrons (marked by two asterisks), that is four particles, four holes. To understand why on top of two-particle, two-hole excitations we also need to add four-particle, four-hole configuration, we list in Table  \ref{tab:amplitudes} the amplitudes of each configuration that has its absolute value greater than 0.1 in the exact calculation. Including two particle-two hole configurations built by excited one neutron and one neutron above the Fermi level already significantly improve the quality of the state. This was already hinted in Fig. \ref{fig:MutualInformation}, where some of the most correlated orbitals are constructed from neutron and proton states. Including two-particle, two-hole operators in the anzatz (\ref{eq:U}) induces higher particle-hole correlations, including four-particles, four-holes. However, if the sum in Eq. (\ref{eq:U}) is truncated to only two-particles, two-holes contributions, one can see immediately in Table \ref{tab:amplitudes} that some of the significant four-particle, four-hole configurations have a much smaller amplitude than what would be required. Explicitly adding four-particle, four-hole configurations significantly improves the quality of the solution, as illustrated in Tables \ref{table:results} and \ref{tab:amplitudes}.

\begin{table*}[t]
    \centering
    \caption{Summary of the results for the ground-state energy for different systems in different model spaces.}
    \begin{tabular}{cccccc}
    \hline \hline
        Nucleus & Model Space & $E_\mathrm{exact}$ (MeV) & $E_\mathrm{HF}$ (MeV) & $E_\mathrm{HF}+\mathrm{corr}$ (MeV) & Number parameters\\
    \hline
    $^6$He & $p$ & $-3.91$ & $-0.90$  & $-3.85$ & 2\\
    $^8$Be & $p$ & $-31.12$ & $-26.12$ & $-26.79$ & 12\\
    $^8$Be$^*$ & $p$ & $-31.12$ & $-26.12$ & $-29.37$ & 76 \\
    $^8$Be$^{**}$ & $p$ & $-31.12$ & $-26.12$ & $-30.67$ & 112 \\
    $^{20}$O & $sd$ & $-23.93$ & $-21.29$ & $-23.18$ & 10 \\
    $^{22}$O & $sd$ & $-35.27$ & $-32.98$ & $-35.14$ & 35 \\
    \hline \hline
    \end{tabular}
    \label{table:results}
\end{table*}

\begin{table}
    \centering
    \caption{Configurations with the absolute value of the amplitude over 0.1 in the exact solutions, compared with approximations of the ground state when two particle, two holes (2p-2h) and four particle, four holes (4p-4h) are introduced on top of the HF solution. The first six digits give the occupation of the neutrons on six states, and the next six, the proton occupation. On the first row we print the amplitude of the HF state.}
    \begin{tabular}{cccc}
\hline \hline
Configuration & Exact & 2p-2h & 4p-4h \\
\hline
000011000011 & $+0.452$ & $+0.760$ &$+0.516$  \\
000011001100 & $-0.201$ & $-0.236$ & $-0.205$ \\
000011110000 & $+0.154$ &  $+0.189$ & $+0.157$ \\
000101000101 & $-0.101$ & $-0.120$ &$-0.104$ \\
001010001010 & $-0.101$ & $-0.120$ & $-0.104$ \\
001100000011 & $-0.201$ & $-0.236$ & $-0.205$ \\
001100001100 & $+0.393$ & $+0.102$ & $+0.386$ \\
001100110000 & $-0.274$ & $-0.068$ & $-0.253$ \\
010100101000 & $-0.146$ & $-0.017$& $-0.135$ \\
011000101000 & $-0.119$ & $-0.020$ & $-0.110$\\ 
100100010100 & $-0.119$ & $-0.020$ & $-0.110$\\ 
101000010100 & $+0.146$ & $+0.017$ & $+0.135$ \\
110000000011 & $-0.154$ &$-0.189$ & $-0.157$\\
110000001100 & $+0.270$ & $+0.067$ &$+0.253$ \\
110000110000 & $-0.239$ & $-0.043$ & $-0.215$\\
    \hline \hline
    \end{tabular}
    \label{tab:amplitudes}
\end{table}

\subsection{Hardware results}

\begin{figure}[t]
    \centering
    \includegraphics{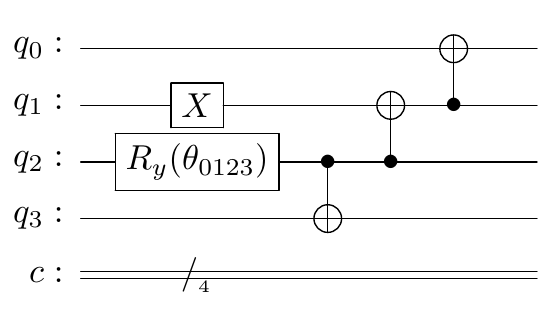}
    \caption{Circuit used to include only the lowest two-particle, two-hole configurations on top of the HF solution. A gate identity was used to adapt the circuit in Fig. \ref{fig:circ-2p2h-JW} upper panel to the connectivity of the \textit{Bogota} machine.}
    \label{fig:circ1d}
\end{figure}

\begin{figure}[h]
    \centering
    \includegraphics[scale=0.32]{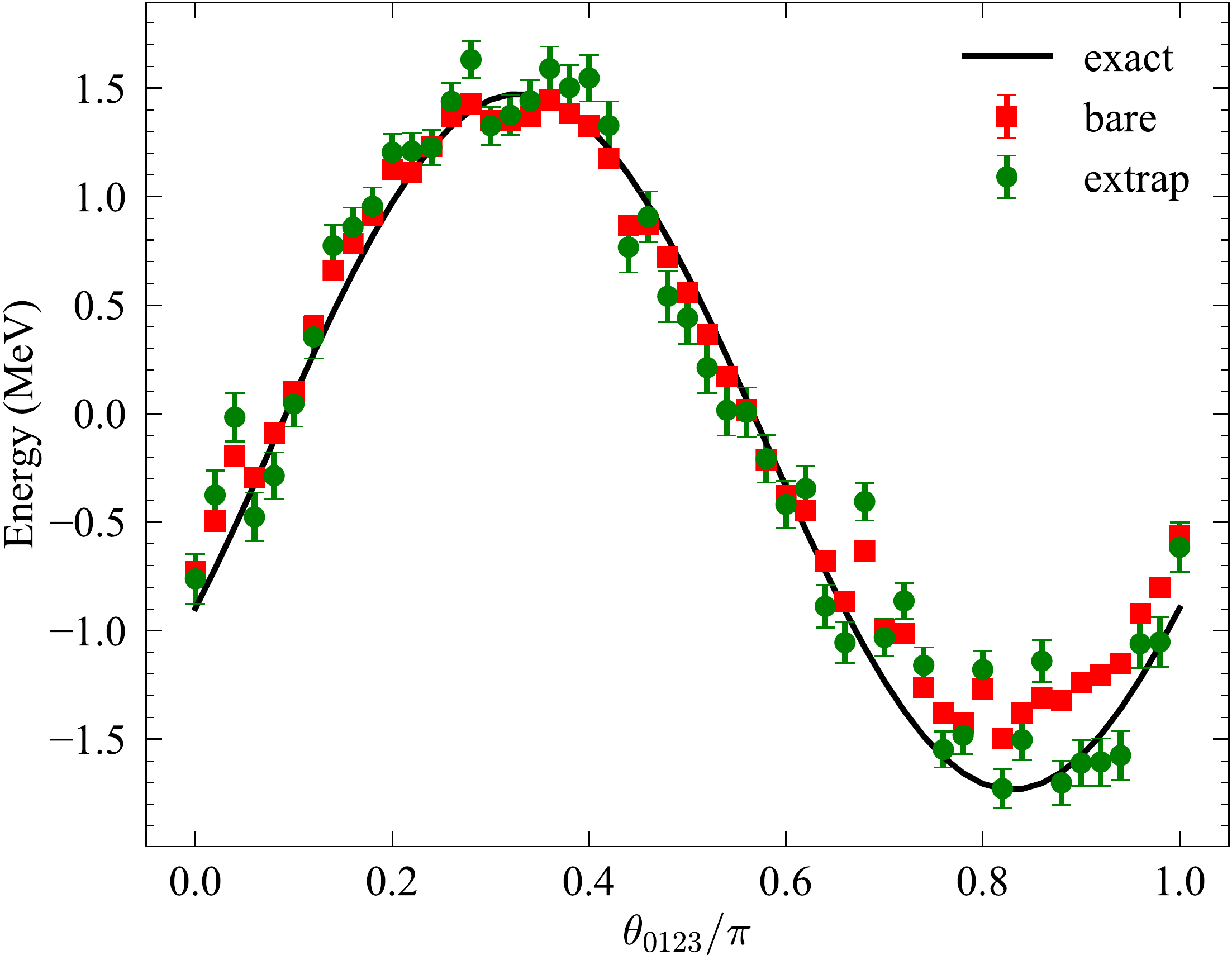}
    \caption{Energy as a function of the angle in the variational ansatz used. We present here the exact and hardware results using the five qubit IBM machine \textit{Bogota}. The red squares show the hardware unmitigated results, while the green circles show our results after readout error mitigation and Richardson extrapolation.}
    \label{fig:plot1d}
\end{figure}

We report here results for runs on noisy simulator and actual hardware. The first Hamiltonian used has $18$ terms and the circuits in the variational ansatz depends on only one parameter and it is reported in Fig.~\ref{fig:circ1d}. We implemented this problem in Qiskit ~\cite{Qiskit21s} and ran it on the IBM quantum device Bogota \cite{IBMQ_Bogota}. A plot of the energy as a function  of the variational parameter $\theta_{0123}$ is reported in Fig.~\ref{fig:plot1d}. We notice that each evaluation of the energy has been error mitigated performing first readout error mitigation using the module included in IBM's Qiskit-Ignis \cite{Qiskit21s} and then mitigated for the noise caused by the CNOT gates. During this last error mitigation procedure for each function evaluation we added $2k$ CNOT gates for each CNOT gate with $k=1,2,3$ and we extrapolated the results to zero noise using Richardson extrapolation as described in Ref.~\cite{Roggero2020c}. We notice that given the fact that we are using very shallow circuits (with a maximum number of $3$ CNOT gates ~\ref{fig:circ1d}) we did not use other methods such as exponential extrapolation, a method that was showed to lead to a significant improvement for observables calculated using significantly higher gate depths ~\cite{Hall2021} than the present ones . 
The error mitigated results are more in agreement with the numerically exact ones, although not in complete agreement.
In order to asses the quality of the results we use the error metric defined in Eq. (33) of Ref.~\cite{Roggero2020c} and reported below for completeness
\begin{eqnarray}
\chi^2&=&\sum_{k=1}^M\frac{\left(v_k^{(e)}-v_{k}^{(t)})^2\right)^2}{\left(\epsilon_k^{(e)}\right)^2}\, ,\\
{\rm nssd}(r)&=&\sqrt{\frac{\sum_{k=1}^M(v_k^{(e)}-v_k^{(t)})^2}{\sum_{k=1}^M(rv_k^{(t)})^2}}\, ,
\end{eqnarray}
where $M$ denotes the number of points used in the angle ($\theta_{0123}$) grid, $v_k^{(t)}$ the exact theoretical result at point $k$ and $v_k^{(e)}$ and $\epsilon_k^{(e)}$ the experimental value and the estimated error respectively. In the following we use $r=0.05$. We recall that $\chi^2$ quantifies the compatibility of the data with the exact results and ${\rm nssd}$ the accuracy of the calculation. The values of both error metrics are reported in Table \ref{tab:err_met}. 
The error mitigation protocol used substantially reduces $\chi^2$. On the other hand, this process leads to a undesirable increase in the sum of squared deviations (nssd) and therefore an increase in uncertainties, and analogue results are obtained using a polynomial fit.

\begin{table}[h]
    \centering
    \caption{ Error metric for the one dimensional problem.}
    \begin{tabular}{cccc}
    \hline \hline
         Type & $\chi^2$& ${\rm nssd}$ \\
         \hline
          Bare &  $2,650$ & $3.6$\\
        Mitigated & $331$ & $4.3$ \\
    \hline \hline
        \end{tabular}
    \label{tab:err_met}
\end{table}

We verified that performing a zero-noise-extrapolation (ZNE) of each one of the terms in the Hamiltonian and then combining them to obtain the full expectation value leads to similar results (and error metrics) of performing directly a ZNE of the full expectation value of the Hamiltonian.  
We also explored a 2D minimization on the virtual machine using the noise model of the IBM five qubit machine Bogota ~\cite{IBMQ_Bogota} using the Hamiltonian obtained from a BK mapping and the ansatz reported in Fig.~\ref{fig:circ-2p2h-JW}, middle panel. We performed some preliminary runs using the new QISKIT feature runtime \cite{Qiskit21s} performing readout error mitigation for each function evaluation and using a variety of minimization methods, obtaining for the best case a value of the energy about $0.5$ MeV higher than the actual ground-state energy. We performed successive runs using the machine noise model applying readout error mitigation and ZNE, in particular using Richardson extrapolation, for each evaluation of the cost function. In this last case the result obtained was almost in agreement with the exact numerical energy as can be seen looking at Table~\ref{tab:BK-results}. It is important to notice that while using the noise model the minimization was successful using only the following derivative free minimizers: COBYLA ~\cite{Powell94}, Powell ~\cite{Powell64} and Nelder Mead ~\cite{Nelder65}. Using stochastic gradient descent based minimizers the results were between $1$ and $2$ (MeV) above the exact ground-state energy. We plan to investigate this empirical observation in future work. We notice here that for realistic nuclear problem sizes the classical minimization procedure should make use of parallelization techniques such as the ones recently developed in Ref.~\cite{Pistoia2020} for gradient descent optimizers. 
%@Ale: please use the Phys Rev style for tables (as below)

\begin{table}[h]
    \centering
    \caption{  Results for the 2D minimization using IBM simulator with the noise model of the five qubit machine Bogota \cite{IBMQ_Bogota}. For the virtual machine run with noise we performed 30 function evaluations. Each function evaluation has been error mitigated following the procedure described in the main text. }
    \label{tab:BK-results}
    \begin{tabular}{cccc}
    \hline \hline
         Type & $\theta_{0123}$& $\theta_{0145}$& $E_{GS}$(MeV)  \\
         \hline
         exact &$2.70 $ & $2.42$ &$ -3.85$\\
          statistical noise &  $2.32$& $   2.30  $& $-3.83 $\\
%          VM bare& & & \\
         VM & $ 2.38 $& $2.61 $& $ -3.73$\\
\hline \hline
          
        \end{tabular}
\end{table}

\iffalse
\begin{table}[h]
    \centering
    \begin{tabular}{c|c|c|c|}
         Type & $\theta_{0123}$& $\theta_{0145}$& $E_{GS}$  \\
         \hline
          statistical noise &  $2.6172$& $0.565$& $-13.838$\\
          \hline
        noise model & $2.475$& $0.745$& $-13.585$\\

        \end{tabular}
    \caption{ L BFGS B method for minimization. The exact ground-state energy obtained diagonalizing the Hamiltonian is $-14.061$ MeV.   }
    \label{tab:1}
\end{table}
\fi
\iffalse
We use the noise model and coupling map of IBMQ casablanca, 8000 shots and Qiskit Ignis readout error mitigation. Minimization is done online using qiskit runtime. The value of the ground-state energies are reported in Table.~\ref{tab:1}

\begin{table}[h]
    \centering
    \begin{tabular}{c|c|c|c|c|}
         method & $E_{GS}$& $\Delta E_{GS}$& $\alpha$ & $\beta$   \\
         \hline
          SPSA &  $-3.378$& $0.070$& $-1.375$& $0.666$\\
         % L BFGS B & $1.013$&$0.067$& $0.771$ & $0.0207$  \\
          COBYLA & $-3.022$ &$0.066$ & $-0.976$ & $0.677$ \\
          Powell & $-3.319$& $0.062$ & $-1.589$ & $0.773$
    \end{tabular}
    \caption{Nelder Mead and  L BFGS B both fail for this problem giving positive energies. The energy is obtained using statevector simulator and $SPSA$ minimization is $-3.854$ with $(\alpha,\beta)=(-1.353,0.708)$. }
    \label{tab:1}
\end{table}
\fi
\iffalse
\begin{itemize}
    \item error mitigation
    \item 1D run
    \item 2D minimization
\end{itemize}
\fi
\section{Summary and Conclusions}

We have investigated the feasibility of present and future quantum computers to prepare the ground state of nuclear systems, with the goal of using quantum hardware in the future to solve nuclear structure and dynamics problems that are too large even for today's leadership-class supercomputers. 
We initially examined various encoding approaches of the many-body problem and estimated their requirements in terms of number of qubits and gate depth.  Depending upon the type of nuclear problem considered, different approaches may be most effective.  For shell-model and related Hamiltonians, second quantized encodings like JW and BK are likely to be most efficient. Of order 50 qubits would be sufficient to perform calculations
beyond what is possible today using classical computers.  However the gate depth requirements (and hence the gate fidelity) are rather strict, with order $10^8$ gates required to prepare a trial state within the UCCSD ansatz. Of course further simplifications could be made to reduce the gate count.
For the case of Hamiltonians with realstic nucleon-nucleon interactions fit to nucleon-nucleon phase shifts,
the number of single particle basis states, typically implemented on a lattice, is considerably larger.
Primarily this is driven by the range of the nucleon-nucleon interaction being comparable to the average
separation between nucleons in a nucleus.  In this case first-quantized methods, which scale only logarithmically in the number of basis, may be optimum, particularly for cases with a modest number of nucleons.

For the second quantized encodings we examined some simple test problems in which only a few nucleons are active in a restricted model space.  In this case exact numerical solutions are available, enabling us to
examine the von Neutman entropy. We also designed circuits for different encodings for these simple problems.
We found that on simulators we can reproduce the exact results, which gives us confidence in the mapping from a second quantization formalism to Pauli strings. However, the hardware quality is not yet fully sufficient to model even relatively simple models like the ones in this study. We did run on both emulators and on actual hardware employing standard error mitigation techniques. These are effective for the most simple cases.
With further advances in quantum hardware and algorithms, we expect a wide variety of problems in nuclear structure and particularly reactions to be amenable to exact solutions.
\iffalse
\appendix
\section{First quantization method}
We report here a derivation of the circuits used in the main text for the first quantized problem. We use a bitonic swap network to generate the anti-symmetrized wave function.
\fi

\section*{Acknowledgments}

We thank Calvin Johnson and Alessandro Roggero for discussions and feedback on the manuscript. IS also thanks Calvin Johnson for providing an updated version of the code \texttt{SHERPA}. This work was carried out under the auspices of the National Nuclear Security Administration of the U.S. Department of Energy at Los Alamos National Laboratory under Contract No. 89233218CNA000001. IS and JC gratefully acknowledge  partial support by the Advanced Simulation and Computing (ASC) Program. AB’s work is supported by the U.S. Department of Energy, Office of Science, National Quantum Information Science Research Centers, Quantum Science Center.

\bibliography{references}

\end{document}